\documentclass[pra,twocolumn,showpacs,preprintnumbers,floatfix]{revtex4-1}
\usepackage{graphicx}% Include figure files
\usepackage{dcolumn}% Align table columns on decimal point
\usepackage{bm}% bold math
\usepackage{latexsym,epsfig}
\usepackage{graphicx}
\usepackage{verbatim}
\usepackage{comment}
\usepackage{amsmath}
\usepackage{stmaryrd}
\usepackage{color}
\usepackage{float}
\usepackage{microtype}

\newcommand{\mrm}{\mathrm}

\newcommand{\vecb}[1]{\mathbf{{#1}}}
\newcommand{\Esca}{\mathcal{E}}

\newcommand{\ef}{\Esca_\mrm{eff}}

\begin{document}
\title{Theoretical Analysis of Effective Electric Fields in Mercury Monohalides}
\author{V.S. Prasannaa$^{1,2}$}
\author{M. Abe$^{3,4}$}
\author{V. M. Bannur$^{2}$}
\author{B.P. Das$^{5}$}

\affiliation{$^{1}$Indian Institute of Astrophysics, Koramangala II block, Bangalore-560034, India}
\affiliation{$^{2}$Department of Physics, Calicut University, Malappuram, Kerala-673 635, India}
\affiliation{$^{3}$Tokyo Metropolitan University, 1-1, Minami-Osawa, Hachioji-city, Tokyo 192-0397, Japan}
\affiliation{$^{4}$JST, CREST, 4-1-8 Honcho, Kawaguchi, Saitama 332-0012, Japan}
\affiliation{$^{5}$International Education and Research Center of Science and Department of Physics,
Tokyo Institute of Technology, 2-12-1-H86 Ookayama, Meguro-ku, Tokyo 152-8550, Japan}

\date{\today}

\begin{abstract}
Mercury monohalides are promising  candidates for electron electric dipole moment searches. This is due to their extremely large values of effective 
electric fields, besides other attractive experimental features. We have elucidated the theoretical reasons of our previous work. 
We have also presented a detailed analysis of our calculations, by including the most important of the correlation effects' contributions. 
We have also analyzed the major contributions to the effective electric field, at the Dirac-
Fock level, and identified those atomic orbitals' mixings that contribute significantly to it. 
\end{abstract}

%\pacs{03.75.Lm, 05.10.Cc, 05.30.Jp} \keywords{Suggested keywords}

\maketitle

\section{Introduction}

The electron electric dipole moment (eEDM) is a consequence of parity and time-reversal 
violations ~\cite{Landau,Ballentine,Sandarsss,Kriplovich}. It is an important non-accelerator probe 
of physics beyond the Standard Model~\cite{Fukuyama,Pospelov}. 
Ibrahim et al make a case that the eEDM can be a sensitive probe of PeV physics~\cite{Ibrahim}. 
There is a large body of work on eEDMs and CP violation in supersymmetric models, for example, see Ref. ~\cite{Rohini}. 
A knowledge of eEDM also 
provides insights into the baryon asymmetry in the Universe (BAU)~\cite{Kazarian,Hisano}. One of the Sakharov conditions~\cite{Sakharov}, 
which gives the necessary prerequisites for 
BAU, is CP violation. 
If the CPT theorem~\cite{CPT} is true, then T violation must correspond to CP violation, to 
preserve CPT symmetry. This correspondence is what connects the two seemingly disparate phenomena, eEDMs and BAU. 
The importance of this connection is demonstrated in the work by Fuyoto et al~\cite{Hisano}, who 
argue that the relationship between the BAU-related CP violations 
and eEDMs are important for the test of the electroweak baryogenesis (EWBG) scenario. They proceed to show that if 
BAU-related CP violation does exist, then the 
EWBG region might be entirely verified by the future eEDM experiments. 

Heavy polar diatomic molecules are currently the preferred candidates to look for a shift in the energy of a 
molecule in a particular state, due to 
the presence of the eEDM (For example, Ref.~\cite{ThO}). 
The electric field corresponding to that shift in energy, with the proportionality constant being the eEDM, 
is called the effective electric field, $E_{eff}$. 
It is the electric field that an electron experiences, due to all other electrons and nuclei in the molecule~\cite{Abe}. 
These calculations warrant a relativistic treatment to compute this quantity, as $E_{eff}$ completely vanishes in the non relativistic
limit~\cite{Sandars}. 

We had calculated the effective electric fields of mercury monohalides and identified them as promising 
candidates for eEDM searches~\cite{VSP}. The main thrust of this work is to elaborate on the theoretical aspects of our previous one~\cite{VSP}. 
In particular, we analyze and elucidate the contributions to the effective electric fields, at the Dirac-Fock and correlation levels.
We employ a relativistic coupled cluster method, for our computations. 

\section{Theory}

The eEDM Hamiltonian, $H_{eEDM}$, is given by

\begin{eqnarray}
 H_{eEDM} = -d_e \sum_{j=1}^{N_e} \beta \vec{\sigma}_j. \vec{E}_{intl,j}
\end{eqnarray}

$d_e$ is the eEDM. The summation is over the number of electrons in the molecule, $N_e$. $\beta$ is one of the Dirac matrices, 
$\vec{\sigma}$ refers to the Pauli matrices, and $\vec{E}_{intl}$ is the internal electric field. 

The shift in energy due to eEDM is given by

\begin{eqnarray}
 \Delta E &=& \langle \psi \arrowvert H_{eEDM} \arrowvert \psi \rangle \\
 &=& -d_e E_{eff}
\end{eqnarray}

Here, $\arrowvert \psi \rangle$ is the ground state wavefunction of mercury monohalides. 
Comparing equations (1) and (3), we obtain the following expression for $E_{eff}$   

\begin{eqnarray}
 E_{eff} &=& \langle \psi \arrowvert \sum_{j=1}^{N_e} \beta \vec{\sigma}_j. \vec{E}_{intl,j} \arrowvert \psi \rangle 
\end{eqnarray}

To obtain the wavefunction, we employ a fully relativistic coupled cluster method (RCCM). The wavefunction is given by 

\begin{eqnarray}
 \arrowvert \psi \rangle = e^T \arrowvert \Phi_0 \rangle
\end{eqnarray}

T is called the cluster operator. $\arrowvert \Phi_0 \rangle$ is the Dirac Fock (DF) wave function. 
We use the relativistic coupled cluster singles and doubles (CCSD) 
approximation in our work. More details about the relativistic CCSD method and its salient features can be found in Ref.s ~\cite{Abe,SrFpaper}. 

The expectation value of any operator, O, in an RCCM, can be expressesd as~\cite{Cizek,Bartlett}: 

\begin{eqnarray}
 \langle O \rangle &=& \frac{\langle \psi \arrowvert O \arrowvert \psi \rangle}{\langle \psi \arrowvert \psi \rangle} \nonumber \\
 &=& \langle \Phi_0 \arrowvert e^{T \dag} O_N e^T \arrowvert \Phi_0 \rangle_C + \langle \Phi_0 \arrowvert O \arrowvert \Phi_0 \rangle
\end{eqnarray}

The subscript, `N' means that the operator is normal ordered~\cite{Lindgren}, and `C' means that each of the terms are connected~\cite{Kvas}. Therefore, 

\begin{eqnarray}
 E_{eff} &=& \langle \Phi_0 \arrowvert e^{T \dag} H^{eff}_{eEDM,N} e^T \arrowvert \Phi_0 \rangle_C + \langle \Phi_0 \arrowvert H^{eff}_{eEDM} \arrowvert \Phi_0 \rangle \\
 &\approx& \langle \Phi_0 \arrowvert (1+T_1+T_2)^{\dag} H^{eff}_{eEDM,N} (1+T_1+T_2) \arrowvert \Phi_0 \rangle_C \nonumber \\ 
 &+& \langle \Phi_0 \arrowvert H^{eff}_{eEDM} \arrowvert \Phi_0 \rangle
\end{eqnarray}

We replace the usual eEDM operator by an effective one~\cite{Abe}, $H^{eff}_{eEDM}$, given by 

\begin{eqnarray}
 \frac{2ic}{e} \sum_{j=1}^{N_e} \beta \gamma_5 p_j^2 
\end{eqnarray}

where 
c is the speed of light, e is the charge of the electron, $N_e$ refers to the number of electrons in the molecule, $\beta$ is one of the Dirac 
matrices, $\gamma_5$ is the product of the Dirac matrices, and $\vecb{p}_j$ is the momentum of the $j^{th}$ electron. This is done, so that 
the Hamiltonian is rewritten in terms of only one-body operators. 
The term $\vec{E}_{intl}$ (from equation (4)) has a two-body operator in it. Although it can be calculated, in principle, 
it is very time demanding and complicated. 
Using an effective one-body operator simplifies the computations by a significant amount. 
Further details can be found in Ref. \cite{Abe}, and the references therein. 
We consider only the linear terms in the expansion of $e^{T}$, both on the bra and the ket sides, in the first term of equation (6), 
as shown in equation (8). 
This is a reasonable approximation, and we can see this from the accuracy of our results from 
our previous works, where we compare them with experimental values~\cite{Abe,SrFpaper,AEM,ASunaga}. 
This approximation, hence, not only saves computational cost by only 
taking into account only the linear terms, but also provides very accurate results. 

Since the dominant contribution to $E_{eff}$ is at the DF level~\cite{VSP}, we analyze the terms that constitute it. The 
contribution, $E_{eff}^{DF}$, can be rewritten as: 

\begin{eqnarray} 
 E_{eff}^{DF} &=& \langle \Phi_0 \arrowvert H_{eEDM}^{eff} \arrowvert \Phi_0 \rangle \nonumber \\
 &=& \sum_i^{MO} \langle \varphi_i \arrowvert h_{eEDM}^{eff} \arrowvert \varphi_i \rangle \nonumber \\
 &=& \langle \varphi_{v} \arrowvert h_{eEDM}^{eff} \arrowvert \varphi_{v} \rangle \nonumber \\
 &=& \frac{4ic}{e} \sum_{k=1}^{NB} \sum_{l=NB+1}^{2NB} C_k^{*S} C_l^L \langle \chi_{v, k}^{S} \arrowvert p^2 \arrowvert \chi_{v, l}^{L} \rangle
\end{eqnarray}

Here, $\varphi_{v}$ refers to the singly occupied molecular orbital (SOMO). $h_{eEDM}^{eff}$ is the single particle effective eEDM operator. 
Summation over the number of molecular orbitals (MO) is indicated by i, while summation over the number of large and small components of the basis sets are 
given by k and l, respectively. NB refers to the number of large component basis funcrtions. $C_k$ and $C_l$ refer to the coefficients, obtained 
by solving the DF equations, and their superscripts L and S stand for large and small components respectively. The $\chi$s refer to the atomic orbitals 
(basis sets) of the constituent atoms. The mixing between large and small components is due to the fact that the eEDM operator is off-diagonal. 
Only the SOMO survives in the expression for $E_{eff}^{DF}$, because the 
remaining terms cancel out. This can be understood in the following way: 

\begin{eqnarray}
 \sum_i^{MO} \langle \varphi_i \arrowvert h_{eEDM}^{eff} \arrowvert \varphi_i \rangle &=& \sum_{i'}^{(MO-1)/2} [ \langle \varphi_{i'} \arrowvert h_{eEDM}^{eff} \arrowvert \varphi_{i'} \rangle \nonumber \\
 &+&  \sum_{\overline{i}'}^{(MO-1)/2} \langle \varphi_{\overline{i}'} \arrowvert h_{eEDM}^{eff} \arrowvert \varphi_{\overline{i}'} \rangle] \nonumber \\
 &+& \langle \varphi_v \arrowvert h_{eEDM}^{eff} \arrowvert \varphi_v \rangle
\end{eqnarray}

In the above expression, we have decomposed the left hand side into three terms. 
The first and the second summation terms on the right hand side denote the contributions from the doubly occupied orbitals in the Kramers pairs, 
$\varphi_{i'}$ and $\varphi_{\overline{i}'}$. 
The third term is the contribution from SOMO. 
The Kramers pair orbitals are related by the time reversal operator ($\tau$)~\cite{Dyallbook}: 

\begin{eqnarray}
 \arrowvert \varphi_{\overline{i}'} \rangle &=& \tau \arrowvert \varphi_{i'} \rangle \\
 - \arrowvert \varphi_{i'} \rangle &=& \tau \arrowvert \varphi_{\overline{i}'} \rangle 
\end{eqnarray}

Therefore, 

\begin{eqnarray}
 \langle \varphi_{\overline{i}'} \arrowvert h_{eEDM}^{eff} \arrowvert \varphi_{\overline{i}'} \rangle &=& \langle \varphi_{i'} \arrowvert \tau^\dag h_{eEDM}^{eff} \tau \arrowvert \varphi_{i'} \rangle \nonumber \\
 &=& - \langle \varphi_{i'} \arrowvert h_{eEDM}^{eff} \arrowvert \varphi_{i'} \rangle
\end{eqnarray}

Hence, the first two terms in equation (11) cancel out pairwise, and only the SOMO remains. 

\section{Results and Discussions}

In this section, we present the method of calculations used in this work, followed by a detailed discussion of the results. 
We used and modified the UTChem code~\cite{UTChem}, for the DF and AO (atomic orbital) to MO 
integral transformations~\cite{aomo}. We performed the CCSD
calculations in the Dirac08 program~\cite{Dirac08}.

The details of the basis sets are given in Table I (uncontracted~\cite{Abe}, 
kinetically balanced~\cite{Dyallbook} Gaussian Type Orbitals (GTOs), cc-pV 
DZ and TZ for F, Cl and Br~\cite{bsl}, Dyall's basis for I~\cite{c2v}, and Dyall's c2v and c3v basis sets for Hg~\cite{c2v}). 
We did not freeze any of our occupied orbitals. We chose the following bond 
lengths (in Angstroms): HgF (2.00686) \cite{knecht}, HgCl (2.42), HgBr (2.62), HgI (2.81) \cite{Cheung1979}. 

\begin{table}[H] 
 \centering
 \begin{tabular}{|c|c|c|}
 \hline
 Atom & Basis (DZ)& Basis(TZ) \\
 \hline
  Hg&22s,19p,12d,9f,1g&29s,24p,15d,11f,2g \\
  F&9s,4p,1d&10s,5p,2d,1f \\ 
  Cl&12s,8p,1d&15s,9p,2d,1f \\ 
  Br&14s,11p,6d&20s,13p,9d,1f \\   
  I&21s,15p,11d&28s,21p,15d \\     
 \hline
 \end{tabular}
 \caption{Summary of the basis sets employed in our calculations. }
\end{table}

\begin{table}[H] 
 \centering
 \begin{tabular}{|c|c|c|c|c|}
 \hline
  Term & HgF & HgCl & HgBr & HgI \\
 \hline
  DF & 104.25 &103.57 &97.89 &96.85 \\
  $H_\mrm{EDM}^\mrm{eff} T_1 +$ cc&20.16 &19.34 &22.18 &24.78 \\
  $T_1^{\dag} H_\mrm{EDM}^\mrm{eff} T_1$ &-3.91 &-3.58 &-4.07 &-4.77 \\
  $T_1^{\dag} H_\mrm{EDM}^\mrm{eff} T_2 +$ cc &0.44 &0.194 &-0.2 &-0.30 \\
  $T_2^{\dag} H_\mrm{EDM}^\mrm{eff} T_2$ &-5.52 &-5.96 &-6.5 &-7.26 \\  
  Total&115.42&113.56&109.29&109.30\\
 \hline
 \end{tabular}
 \label{tab:contributions}
 \caption{Contributions, from the individual terms, to the effective electric field of HgX. cc refers to complex conjugate, of the term that it 
 accompanies. }
\end{table}

Table II shows the terms from equation (8), 
and also the total $E_{eff}$, for HgX. In our previous work, we had performed this analysis for only HgF. Extending it to all HgX 
enables us to study the trends in $E_{eff}$, 
for these molecules. Also, in our previous work, we had identified HgBr as a promising candidate, among the HgX molecules. Hence, theoretical details about
HgX other than HgF become important. 
In the notation used in the Table, $H_\mrm{eEDM}^\mrm{eff} T_1$, for example, actually refers to 
$\langle \Phi_0 \arrowvert \{H_\mrm{eEDM}^\mrm{eff}\} T_1 \arrowvert \Phi_0 \rangle_C$, where the curly brackets refer to a normal-ordered 
operator. This is done for the purpose of brevity. 
Note that the $\langle \Phi_0 \arrowvert \{H_\mrm{eEDM}^\mrm{eff}\} \arrowvert \Phi_0 \rangle_C$ is zero, since the effective eEDM operator is normal 
ordered~\cite{AEM}. 
The $H_\mrm{eEDM}^\mrm{eff} T_2$ term, and its complex 
conjugate, are zero, due to the Slater-Condon rules~\cite{Lindgren}. Hence, we are left with seven non-zero terms. 

The DF term is the largest, and it decreases from HgF to HgI. 
Correlation effects account for 
about 9 percent of the total effective field. 
This indicates that for these molecules, both $E_{eff}$ and the amount of correlation does not significantly vary with Z of the lighter, halide atom. 
Among the correlation terms, the $H_\mrm{eEDM}^\mrm{eff} T_1$ term is the largest. The second and the third largest 
correlation contributions come from the $T_2^{\dag} H_\mrm{eEDM}^\mrm{eff} T_2$ term, and $T_1^{\dag} H_\mrm{eEDM}^\mrm{eff} T_1$ term. Their effect is to 
reduce $E_{eff}$. 
The overall values of $E_{eff}$ decrease from HgF to HgBr. HgBr and HgI have almost the same values of $E_{eff}$, although the DF value of 
HgI is smaller in comparison with HgBr. 
This can be understood from the fact that the difference between the $H_\mrm{eEDM}^\mrm{eff} T_1$ term, and the 
$T_1^{\dag} H_\mrm{eEDM}^\mrm{eff} T_1 + T_2^{\dag} H_\mrm{eEDM}^\mrm{eff} T_2$ terms is larger for HgI. 

We shall remark briefly about how the correlation trends vary in the $E_{eff}$s of HgX, as compared to those in our previous and ongoing works. 
In our previous work on the PDM of SrF~\cite{SrFpaper}, and subsequently on the PDMs of the other alkaline earth monofluorides (AEMs)~\cite{AEM}, 
we had performed the same analysis. Although $E_{eff}$ and PDM are different properties, they do share similarities, for example, 
both the properties depend on the mixing of orbitals of opposite parity. Hence, it is worthwhile to check if there are any similarities in 
their correlation trends. We first compare the correlation trends between the $E_{eff}$ of HgX and the PDMs of the AEMs. 
Both of them are systems with one unpaired electron, but we see that in AEMs, while correlation 
can decrease (for example, BeF, by around 20 percent) or increase (for example, BaF, by around 20 percent) the PDM, the effect of 
correlation on the $E_{eff}$s of HgX is almost the same throughout, from HgF to HgI. The PDMs of HgX follow an entirely different trend, 
where the correlations decrease the PDM drastically~\cite{VSPpdm}. 
We now compare the correlations in the $E_{eff}$s of HgX with those in YbF, the candidate that currently sets the second best limit on eEDM, and 
RaF, a promising candidate for eEDM experiments. In all the HgX molecules, correlations account for about ten percent. In YbF, the correlations account for 
about twenty percent~\cite{Abe}, while in RaF~\cite{unpublished}, it is close to thirty percent. 
Again, all of these systems have one unpaired electron, and their heavier 
atoms have atomic numbers fairly close to one another, but their effective fields and their correlation effects are very different. 
In HgI, for example, the correlation effects are ten percent, owing to the fact 
that nearly half of the $H_\mrm{eEDM}^\mrm{eff} T_1$ term is cancelled out by the other correlation terms. 
In RaF, this is not so. In fact, the $H_\mrm{eEDM}^\mrm{eff} T_1$ term adds to about 20 GV/cm, and the rest close to -0.5 GV/cm. 
The values that we finally obtain are a consequence of several cancellations at 
work, among the various DF and the correlation terms. We shall attempt to understand further these cancellations for HgX in the rest of this paper. 
This brief discussion illustrates that further detailed theoretical studies are required to understand better the correlation effects and trends, 
of these class of polar molecules. 

Figure I shows some of the dominant Goldstone diagrams involved in the expectation value expression, given by equation (8), 
specifically the DF, the $H_\mrm{eEDM}^\mrm{eff} T_1$, the $T_1^{\dag} H_\mrm{eEDM}^\mrm{eff} T_1$ term, the 
direct counterparts of the $T_1^{\dag} H_\mrm{EDM}^\mrm{eff} T_2$, and the 
$T_2^{\dag} H_\mrm{eEDM}^\mrm{eff} T_2$ terms. The conjugate diagrams are not given, since they give the same result. 

\begin{figure}[H]% t->top of the page, b->bottom of the page, h->'here', as placed in the tex
   \centering
\psfig{file=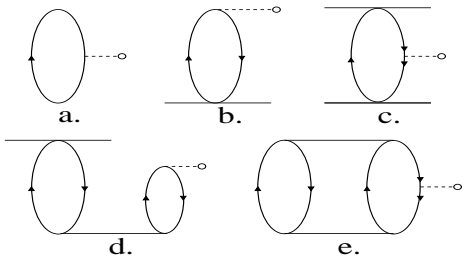,height=1.3in,width=2.4in,angle=0}
\caption{Goldstone diagrams for $E_{eff}$: a. DF term, b. $H_\mrm{eEDM}^\mrm{eff} T_1$, c. the $T_1^{\dag} H_\mrm{eEDM}^\mrm{eff} T_1$ term,  
d. Direct diagrams of the $T_1^{\dag} H_\mrm{eEDM}^\mrm{eff} T_2$ term and the $T_2^{\dag} H_\mrm{eEDM}^\mrm{eff} T_2$ term, respectively. }
\label{figure:Figure 1}%you can label figure as anything you want (here it is 'fig1'). in the text while referencing this figure
% just write Fig.(\ref{fig:fig1}). samething for the equations, label equations as \label{eq:eqn1} and refer it in text like figures
\end{figure}
 
The physical interpretation of these diagrams are 
discussed in detail in another work on PDMs~\cite{AEM}. For the sake of completeness, we choose a representative diagram, $H_\mrm{eEDM}^\mrm{eff} T_1$, to 
explain its physical significance. We choose this diagram, since it contributes the most to the effective electric field. 
This term can be expanded as 

\begin{eqnarray}
 \sum_{i,a} [\langle \varphi_i \arrowvert \{h_\mrm{eEDM}^\mrm{eff}\} \arrowvert \varphi_a \rangle \langle \varphi_a \arrowvert t \arrowvert \varphi_i \rangle]_C
\end{eqnarray}

where the summation is over i and a, where i refers to the occupied orbitals (holes), and a to the virtual orbitals (particles). 
We obtain this expression, if we apply the Slater-Condon rules to the original expression, 
$\langle \Phi_0 \arrowvert \{H_\mrm{eEDM}^\mrm{eff}\} T_1 \arrowvert \Phi_0 \rangle_C$. 
Mathematically, the $H_\mrm{eEDM}^\mrm{eff} T_1$ term represents an all-order residual Coulomb interaction, resulting in an electron 
from an occupied orbital, i, being excited to a virtual, a, and then  
falling back into the same state, i, due to the interaction of the particle with the eEDM. 
This diagram represents several correlation effects, like the Brueckner pair correlation (BPC)~\cite{Lindgren}, 
among others, but is 
not obvious from the coupled cluster diagram, since the $T_1$ part embodies in it the residual Coulomb interaction, to 
all orders of perturbation.  

Table III presents the various contributions to the DF value of $E_{eff}$, due to the mixing between various orbitals 
(or basis sets) (equation (10)), at the TZ level. We have not presented the analysis for the DZ basis sets, since TZ is closer to 
the actual wave function, and the results from both the basis sets show the same trend. 
In the second column, $s-p_{1/2}$, for example, is actually a shorthand for  
$\sum_{s} \sum_{p_{1/2}} C_{s}^{*S} C_{p_{1/2}}^L \langle \chi_{v, s}^{S} \arrowvert p^2 \arrowvert \chi_{v, p_{1/2}}^{L} \rangle$. 
The first summation is over all the small component basis sets of the s angular momentum function, and the second over the large component basis sets 
of the $p_{1/2}$ angular momentum function. 
The mixing between the same parity 
orbitals is zero, and hence, those terms that contain matrix elements between s and d, for example, are ruled out. Only s, $p_{1/2}$, 
$p_{3/2}$, $d_{3/2}$, $d_{5/2}$, and $f_{5/2}$ orbitals for Hg, and s and $p_{1/2}$ for X, have been considered in equation (10), since the 
terms involving the mixing between 
other orbitals contribute negligibly to $E_{eff}^{DF}$. This can also be recognized from the difference between the rows labelled, `Total', which 
gives the sum of the mixings associated with the orbitals considered, and `DF', which gives the total DF contribution. The difference 
between the two decrease from F to Br. 

The combined $s-p_{1/2}$ and $p_{1/2}-s$ contribution is clearly the highest among all others, 
contributing to over hundred percent of the total DF value of $E_{eff}$ in all cases, except HgI. 
The `anomaly' in HgI is due to the halide atom's contribution becoming important. 

We observe that the absolute magnitude of all the terms for Hg decreases from F to I. 
However, in X, we see the opposite trend. In fact, for HgI, the contribution from X increases the effective field by over 2.5 GV/cm. 

The angular momentum functions are strictly not atomic orbitals, but the terms from the basis sets that we employ. 
Hence, we cannot, from our results alone, deduce those principal quantum numbers that contribute significantly. 
However, we can intuitively guess that the major contribution 
is from the 6s and the virtual $6p_{1/2}$ orbitals of the Hg atom, since they lie close in energy, and their radial overlap is large. Moreover, 
we can expect the matrix elements of the eEDM operator between these opposite parity orbitals to be large. 

The importance of $s-p_{1/2}$ and $p_{1/2}-s$ mixing of the heavier atom in the $E_{eff}$ of HgF 
has been understood in the past, for example, Ref.s~\cite{Meyer}. 
We shall compare our results with the previous ones later in this manuscript. 
In our work, we have taken into account not only the $s-p_{1/2}$ and $p_{1/2}-s$ mixing, 
but also that of the other orbitals of both the atoms, and then demonstrate that 
it is the $s-p_{1/2}$ mixing of the heavier atom that dominates.
In the Table, we have only shown the $s-p_{1/2}$ and $p_{1/2}-s$ mixing of the lighter atom, but that is because we 
found the other mixings to be negligibly small. Also, note that our analysis is not only for HgF, but for the heavier mercury monohalides too. 
For example, in HgI, the `lighter' atom, iodine, is sufficiently heavy. 
In spite of that, we see that the $s-p_{1/2}$ and $p_{1/2}-s$ mixing from I is surprisingly small. 

Finally, we observe that not only the magnitude of the $s-p_{1/2}$ and the $p_{1/2}-s$ mixings (of the heavier atom) 
are large, but so is the remainder when these terms 
cancel each other's contributions. This illustrates the importance of cancellations that occur in ab initio calculations. In the case of iodine (in HgI), 
the two terms themselves are non-negligible, but they cancel each other out, leaving behind a very small contribution to the DF $E_{eff}$ from the `lighter' 
atom. 

\begin{table}[H] 
 \centering
 \begin{tabular}{|r|r|r|r|r|r|}
 \hline
 Atom&Mixing &HgF&HgCl&HgBr&HgI  \\
 \hline
	    
Hg&      $s-p_{1/2}$	   &  	     -266.29&	-262.07	 &   -249.39&	-242.34	\\    
Hg&      $p_{1/2}-s$	   &         373.37&	367.74	  &   349.42&	339.56	  \\   
Hg&      $p_{3/2}-d_{3/2}$&          31.22&	25.22	  &    21.84&	20.99	\\     
Hg&      $d_{3/2}-p_{3/2}$ &        -32.26&	-26.35	  &   -22.48&	-21.84	\\      
Hg&      $d_{5/2}-f_{5/2}$&	    -0.91&	-0.51	  &    -0.39&	-0.33	\\      
Hg&      $f_{5/2}-d_{5/2}$&	   0.92	  &       0.52	  &     0.4&	0.33	\\       
X &      $s-p_{1/2}$	   &     -2.78	   &     -4.85	  &   -10.58&	-17.19	\\      
X &      $p_{1/2}-s$	    &     2.79	    &    4.92	  &     11.17&	19.87	\\       
Total: &&	          106.06  &     104.62	  &     99.99&	99.05	\\   
DF&&105.47&104.03&99.55&98.99\\
&$s-p_{1/2}$ and $p_{1/2}-s$&     107.08	  &   105.67	  &    100.03&	97.22	\\

 \hline
 \end{tabular}
 \caption{Summary of the DF results, of the contributions from various orbitals' mixings, at the TZ level}\label{table:t3}
\end{table}

%We shall now look at the expressions for each of the non-zero terms of equation (11). We ignore the expressions for the complex conjugate counterparts. 
%Also, we are not explicitly mentioning that each of the terms are connected, or that the effective eEDM operator is normal ordered. 

We shall now attempt to explain why the $H_\mrm{eEDM}^\mrm{eff} T_1$ term is large, among the correlation terms. 
The DF contribution dominates among the others, due to the significantly high difference 
between the large values of $s-p_{1/2}$ and 
$p_{1/2}-s$ of the Hg atom (the notation is same as that in Table III). 
We wish to reiterate that s is an occupied orbial, while $p_{1/2}$ is a virtual. 
Now, let us focus only on the matrix elements. Their values are several orders larger than their corresponding coefficients. 
These large matrix elements, of the form,$ \langle o \arrowvert h_\mrm{eEDM}^\mrm{eff} \arrowvert v \rangle$ 
(where we abbreviate occupied orbitals as, `o', and virtual as, `v'), also occur 
in the expression for $H_\mrm{eEDM}^\mrm{eff} T_1$ (and hence it also contains matrix elements between s and $p_{1/2}$), except 
that $H_\mrm{eEDM}^\mrm{eff} T_1$ has accompanying it a $t_1$ amplitude, which indicates the ``weight" associated with a one-hole 
one-particle excitation. 
The t amplitudes are like probability amplitudes, and thereby lesser than one always. 
Hence, we can view the amplitude as having a ``reducing" effect on the $H_\mrm{eEDM}^\mrm{eff} T_1$ term, for each i and a. This is probably why 
the $H_\mrm{eEDM}^\mrm{eff} T_1$ term is not as large as the DF one; the large matrix elements are accompanied by the smaller values 
of the $t_1$ amplitudes. Obviously, this is not the only reason why the DF term is substantially larger than the other terms. 
The term also contains matrix elements of the type 
$ \langle o \arrowvert h_\mrm{eEDM}^\mrm{eff} \arrowvert o \rangle$ and 
$ \langle v \arrowvert h_\mrm{eEDM}^\mrm{eff} \arrowvert v \rangle$, which  
cancel each other in a way that results in the final value of the DF term.  There are also cancellations 
between various terms that constitute $H_\mrm{eEDM}^\mrm{eff} T_1$, since not all matrix elements or the $t_1$ amplitudes are 
positive. 

The matrix elements, of the form $ \langle o \arrowvert h_\mrm{eEDM}^\mrm{eff} \arrowvert v \rangle$, occur only 
in $H_\mrm{eEDM}^\mrm{eff} T_1$ and $T_2^{\dag} H_\mrm{eEDM}^\mrm{eff} T_1$ (and its complex conjugate term). The latter, 
however, is not as large as the former, and is, in fact, very small, probably due to two reasons. 
The first is the cancellations that arise among the four terms that constitute $T_2^{\dag} H_\mrm{eEDM}^\mrm{eff} T_1$: 

\begin{eqnarray}
 \sum_{i>j,a>b} [t_{ij}^{ab*}t_i^a \langle b \arrowvert h_\mrm{eEDM}^\mrm{eff} \arrowvert j \rangle + t_{ij}^{ab*}t_j^b \langle a \arrowvert h_\mrm{eEDM}^\mrm{eff} \arrowvert i \rangle \nonumber \\
 -t_{ij}^{ab*}t_j^a \langle b \arrowvert h_\mrm{eEDM}^\mrm{eff} \arrowvert i \rangle - t_{ij}^{ab*}t_i^b \langle a \arrowvert h_\mrm{eEDM}^\mrm{eff} \arrowvert j \rangle]
\end{eqnarray}

We have not explicitly mentioned that the operator is normal ordered, or that each term is connected. 
The four terms are 0.72, 0.24, -0.03, and 0.77 respectively, for HgF, which we choose as a representative case (we expect the trends to 
be similar for the other monohalides). 
All the four terms have matrix elements of the form, $\langle o \arrowvert h_\mrm{eEDM}^\mrm{eff} \arrowvert v \rangle$, of 
which two are dominant, and they almost cancel each other out. 
The second reason is that in this term, there is another t amplitude, $t_2$, which is also less than one, and hence ``reduces" the 
net contribution further. 

The $T_1^{\dag} H_\mrm{eEDM}^\mrm{eff} T_1$ consists of terms of the type $\langle o \arrowvert h_\mrm{eEDM}^\mrm{eff} \arrowvert o \rangle$ and 
$\langle v \arrowvert h_\mrm{eEDM}^\mrm{eff} \arrowvert v \rangle$. To discern how these terms contribute, 
we expand the $T_1^{\dag} H_\mrm{eEDM}^\mrm{eff} T_1$ 
term: 

\begin{eqnarray}
 &-& \sum_{i,j,a}t_i^{a*}t_j^a \langle j \arrowvert h_\mrm{eEDM}^\mrm{eff} \arrowvert i \rangle + \sum_{i,a,b} t_i^{a*}t_i^b \langle a \arrowvert h_\mrm{eEDM}^\mrm{eff} \arrowvert b \rangle \nonumber \\
 &-& \sum_{i,a}t_i^{a*}t_i^a \langle i \arrowvert h_\mrm{eEDM}^\mrm{eff} \arrowvert i \rangle + \sum_{i,a} t_i^{a*}t_i^a \langle a \arrowvert h_\mrm{eEDM}^\mrm{eff} \arrowvert a \rangle \\
 &=& I + II + III + IV
\end{eqnarray}

The first term corresponds to Figure 1. c. The second term is similar to the figure, except that the eEDM vertex is between two particles, instead of 
two holes. The third term is same as the first, except that in its Goldstone diagram, both the holes are the same orbitals, that is, 
the interaction of the hole with the eEDM leaves it unchanged. The fourth is again, same 
as the second, but with the two particles being the same orbital. We have expanded $T_1^{\dag} H_\mrm{eEDM}^\mrm{eff} T_1$ this way, so that we can 
understand which type of matrix elements contribute to it dominantly. Table IV summarizes the contributions to this term. 

\begin{table}[H] 
 \centering
 \begin{tabular}{|l|c|}
 \hline
  Term & Contribution (GV/cm) \\
 \hline
   I&$~10^{-3}$\\
  II&2.94\\
  III&-4.44\\
  IV&2.4\\
 \hline
 \end{tabular}
  \caption{The terms that contribute to $T_1^{\dag} H_\mrm{eEDM}^\mrm{eff} T_1$}
 \end{table}

We observe from the Table that the magnitude-wise, the terms that contain the eEDM operator between the same 
orbitals (terms II and III) contribute significantly. 
Note that these matrix elements are non-zero, although $h_\mrm{eEDM}^\mrm{eff}$ is P-odd. This is because the orbitals are MOs, and each MO is expanded as a 
linear combination of basis functions, of different angular momenta. The major contributions to the $T_2^{\dag} H_\mrm{eEDM}^\mrm{eff} T_2$ term is also from 
matrix elements between the same orbitals (-4.7 GV/cm). 
 
Table V summarizes the results obtained from previous works. Only results for HgF are available, and we proceed to briefly discuss them. 
The first work on the $E_{eff}$ of HgF was by Kozlov~\cite{Dmit2}. 
It was a relativistic, 
semi-empirical calculation. The focus of the paper was the nuclear anapole moment, and electron-nucleus P and T violating 
interactions. The table of results gives the final result of $E_{eff}$. Note that since it is a semi-empirical calculation, it cannot 
break the final value of $E_{eff}$ into its constituent DF and correlation parts. 

Dmitriev et al~\cite{Dmitriev1992} computed the $E_{eff}$ of HgF, using their calculated bond length of 2.11 Angstrom. 
They chose the minimal atomic basis set for F, while for Hg, they 
used five relativistic valence orbitals, $5d_{3/2}$, $5d_{5/2}$, $6s_{1/2}$. $6p_{1/2}$, and $6p_{3/2}$. They obtained a value of about 100 GV/cm. 
Their calculation can be described as quasi relativistic, since it requires the addition of the spin-orbit interaction to a non-relativistic Hamiltonian. 
Our work is fully relativistic (We do not resort to an effective Hamiltonian, but the Dirac-Coulomb one) and has 
the spin-orbit interaction and other effects built into it. 
They did not account for mixing between orbitals beyond $d_{5/2}$, the effect of 
F was ignored, and also only the principal quantum numbers 5 and 6 were chosen. 
We have made no such restrictions. Finally, they had adopted CI, exciting only three
outer electrons. In our CCSD calculation, all the 89 electrons were excited. Hence, the Hilbert space that we 
considered to capture the correlation effects is 
larger than that in their work. Our error estimate of 5 percent is 
better than their estimate of 20 percent. Their estimate of $E_{eff}$ does not contain in it information on the DF or correlation 
contributions. We have provided a detailed 
breakdown of $E_{eff}$ in our work. 
Also, their close agreement with our results for $E_{eff}$ may be a result of fortuitous cancellations. 
For example, their work computeed the PDM of HgF to 
be 4.15 D, which is close to our DF value of 3.9 D. But, our relativistic CCSD result is 2.61 D! 

Meyer et al~\cite{Meyer} calculated $E_{eff}$ for HgF, among 
several other molecules, using their non-relativistic software, to compare the accuracy of their mehod. Later, in 2008, they improved upon their 
approach further, to obtain a more accurate value, of 95 GV/cm~\cite{Meyer2}. 

In our previous work on HgX~\cite{VSP}, we had taken recourse to the relativistic 
coupled cluster method. We had shown that $E_{eff}$,  for all the HgX molecules, is substantially larger than that for all the 
current eEDM molecular candidates. However, we had
not performed a detailed analysis of the physical effects at the DF and correlation levels, which is what we have done in the present manuscript. 
We wish to amphasize that besides the fact that we use a fully relativistic coupled cluster approach and extend the results to all HgX, we also 
break the final value of $E_{eff}$ down into its constituent terms, both at the DF and correlation levels. 

\begin{table}[H] 
 \centering
 \begin{tabular}{|l|c|}
 \hline
  Work & $\ef$ (GV/cm) \\
 \hline
  Y Y Dmitriev \emph{et al.} \cite{Dmitriev1992} & 99.26 \\
  Meyer \emph{et al.}\cite{Meyer} & 68 \\
  Meyer \emph{et al.}\cite{Meyer2} & 95 \\
  This work & 115.42 \\ 
 \hline
 \end{tabular}
  \caption{Effective electric field, $\ef$, in the HgF molecule, calculated in earlier literature}
\end{table} 

Since we are elaborating on the theoretical aspects of our previous work, 
the error estimates are the same (Refer ~\cite{VSP}). 
We recently improved upon our earlier results on HgX $E_{eff}$s, where we take into account the effect of the neglected non-linear 
coupled cluster terms in the expectation value~\cite{ffcc}. Therefore, the non-linear terms, in fact, 
contribute far lesser than our earlier estimate of 3.5 percent. 

\section{Conclusion}

We have calculated the effective electric fields of mercury monohalides. 
We have not frozen any of the core orbitals. We employed Dyall's basis sets for Hg and I, and cc-pV basis sets for the other halides. 
The DF term contributes the most to $E_{eff}$ (about ninety percent). 
We have reported the trends 
in some of the correlation terms for these molecules, at the DZ level. We observe that the dominant contribution to the correlation 
effects is from a one hole-one particle excitation coupled cluster diagram. We present one example of a physical effect that is included in this 
diagram. We have also reported on those mixings of atomic orbitals that significantly 
contribute to the DF value of $E_{eff}$, and observed their trends, at the TZ level. We recognize that the $s-p_{1/2}$ mixing in Hg contributes 
dominantly to $E_{eff}^{DF}$. 

\section{Acknowledgment}

Computations were performed on the GPC supercomputer at the SciNet HPC Consortium. SciNet is funded by: the Canada Foundation 
for Innovation under the auspices of Compute Canada; the Government of Ontario; Ontario Research Fund - Research Excellence; and the 
University of Toronto~\cite{Scinet}. 
We also used the Hydra cluster, in IIA. 
This research was supported by
JST, CREST. M.A. thanks MEXT for financial support. 
The DiRef database was extremely useful in 
searching for literature~\cite{diref}.

\end{document}